\documentclass[12pt]{iopart}
\usepackage{amssymb,graphicx}
\everymath{\displaystyle}
\begin{document}
\newcommand{\Vee}{\mathop{\vee}}
\newcommand{\Wedge}{\mathop{\wedge}}
\title{Initial value problem of evolution equations defined by lattice operators}
\author{Takatoshi Ikegami$^1$, Daisuke Takahashi$^1$, Junta Matsukidaira$^2$}
\address{$^1$ Major in Pure and Applied Mathematics, 3-4-1, Okubo, Shinjuku-ku, Tokyo 169-8555, Japan}
\address{$^2$ Department of Applied Mathematics and Informatics, Ryukoku University, 1-5, Yokotani, Seta Oe-cho, Otsu, Shiga 520-2194, Japan}
\ead{daisuket@waseda.jp}
\begin{abstract}
We propose dynamical systems defined on algebra of lattices, which we call `lattice equations'.  We give exact general solutions of initial value problems for a class of lattice equations, and evaluate the complexity of the solutions.  Moreover we discuss the relationship between those equations and binary cellular automata.
\end{abstract}
\pacs{02.30.Ik, 05.45.-a, 45.05.+x, 45.50.-j}
\maketitle
\section{Introduction}
  Lattices are basic algebra that provide a natural way to formalize and study the ordering of objects.  They play an important role in various fields such as mathematics, computer science and engineering\cite{gratzer}.  For example, they have applications in combinatorics, number theory, group theory, graph theory, set theory and logic circuits.\par
  Lattices have binary operations on elements of (partially) ordered sets called join ($\Vee$) and meet ($\Wedge$).  By these operations, we can investigate structure of ordered sets and manipulate elements of them algebraically.  An application of Boolean lattice to logic circuits is one of successful case of such manipulation.\par
  Meanwhile, there have been numbers of researches on complex dynamical systems that are described by differential equations or difference equations.  Most of complex dynamical systems exhibit either stable behavior, periodic behavior, or unpredictable behavior called chaos, depending on initial values and parameter of the systems.  To classify the behavior of the systems, several entropy-based complexity measures have been proposed and employed\cite{bellon,grammaticos}.\par
  In this article, we propose dynamical systems defined on algebra of lattices, which we call `lattice equations'.  We give exact general solutions of initial value problems for a class of lattice equations, and evaluate the complexity of the solutions.  Since lattice equations are reduced to a certain class of max-plus equations or binary cellular automata (CA)\cite{gaubert,wolfram,wolfram2} when we restrict dependent variables to the ordered set $(\mathbb{R},\le)$ or binary numbers, exact solutions of them apply to max-plus equations and CA.  Though numerical or statistical studies about the behavior of solutions have been vastly done for max-plus equations and CA, direct evaluation of behavior of solution by means of explicit expressions is quite few.  Therefore, our approach provides a novel viewpoint to analysis of complex dynamical systems, that is, analytical mathematics on lattice equations.\par
  We note our result is related to ultradiscretization procedure in the field of integrable systems that connects discrete max-plus equations, CA, and continuous differential or differences equations\cite{takahashi,tokihiro,nishinari,takahashi2}.  The solutions are also directly connected one another through the procedure, though it is mainly found for the integrable systems.  In this context, the lattice equations proposed with exact solutions will be useful in finding wider perspective between continuous and digital systems.\par
  The contents of this article are as follows.  In Section~\ref{sec:lattice}, we first introduce the general definition of lattice and define a class where the complexity of their solutions are of polynomial order.  Secondly we introduce the definition of elementary cellular automata (ECA) and show the relation between lattice equations and ECA.  In Section~\ref{sec:solution}, we show a list of lattice equations of the above class together with their exact solutions.  In Section~\ref{sec:asymptotic}, we show some examples analyzing the asymptotic behavior of ECA using the exact solutions.  In Section~\ref{sec:conclusion}, we give concluding remarks.  In \ref{sec:list}, we give a list of the lattice equations and corresponding ECA which are equivalent to each other under the transformation of variables.  In \ref{sec:proof}, we give some proofs about the solutions listed in Section~\ref{sec:solution}.
\section{Lattice and Elementary Cellular Automaton}  \label{sec:lattice}
\subsection{Lattice}  \label{subsec:lattice}
A partially ordered set (poset) $(L,\le)$ is a lattice if the supremum (join) and the infimum (meet) of $\{a,b\}$ always exist for any $a$, $b\in L$\cite{gratzer}.  Let us define $\vee$ and $\wedge$ by
\begin{equation}
  a\vee b=\sup\{a,b\}, \qquad a\wedge b=\inf\{a,b\}.
\end{equation}
We have the following laws for any lattice,
\begin{eqnarray}
  a\vee b=b\vee a, \quad a\wedge b=b\wedge a, & \mbox{(commutative laws)} \\
  a\vee(b\vee c)=(a\vee b)\vee c,\\
  a\wedge(b\wedge c)=(a\wedge b)\wedge c,
   & \mbox{(associative laws)} \\
  a\vee(a\wedge b)=a, \quad a\wedge(a\vee b)=a, \quad & \mbox{(absorption laws)} \\
  a\vee a=a, \quad a\wedge a=a. & \mbox{(idempotent laws)}
\end{eqnarray}
Note that we can define $\vee$ and $\wedge$ by
\begin{eqnarray}
  a\vee b=\max(a,b), \qquad a\wedge b=\min(a,b),
\end{eqnarray}
for the totally ordered set.\par
The poset $(L,\le)$ is a `distributive lattice' if it satisfies (one of) the following equivalent distributive laws for any $a$, $b$ and $c\in L$,
\begin{eqnarray}
  & a\vee(b\wedge c)=(a\vee b)\wedge(a\vee c), \nonumber\\
  & a\wedge(b\vee c)=(a\wedge b)\vee(a\wedge c), \qquad\qquad\qquad\mbox{(distributive laws)}\\
  & (a\vee b)\wedge(b\vee c)\wedge(c\vee a)=(a\wedge b)\vee(b\wedge c)\vee(c\wedge a). \nonumber
\end{eqnarray}
For example, the totally ordered set $(\mathbb{R},\le)$ for the real number field $\mathbb{R}$ with the usual order $\le$ is a distributive lattice since the above distributive laws hold.\par
  Let us consider a `conjugate' element $\overline{a}$ for each $a\in L$ by the conditions,
\begin{equation}
  \overline{a}\in L, \quad \overline{\overline{a}}=a, \quad \overline{a}\ge\overline{b}\mbox{\quad if $a\le b$}.
\end{equation}
Note that conjugate elements can not be defined for a general lattice and the definition is not unique even if they can be.  For example, we can define the conjugate element for $(\mathbb{R},\le)$ by
\begin{equation}
  \overline{a}=c-a\quad (\mbox{$c$: constant})
\end{equation}
and also by
\begin{equation}
  \overline{a}=\cases{ -ca & ($a\le 0$) \\ -a/c & ($a>0$)},
\end{equation}
where $c>0$.  For the conjugation, the following laws always hold,
\begin{equation}
  \overline{a\vee b}=\overline{a}\wedge\overline{b},
\qquad
  \overline{a\wedge b}=\overline{a}\vee\overline{b}.
\end{equation}
\subsection{Evolution equations expressed by lattice operations}  \label{subsec:evolution}
Let us consider the following form of evolution equations using lattice operations,
\begin{equation}  \label{lattice eq}
  u_j^{n+1}=f(u_{j-1}^n,u_j^n,u_{j+1}^n),
\end{equation}
where $j$ denotes an integer space site number, $n$ an integer time and $u\in L$.  The function $f(a,b,c)$ is constructed by $\vee$, $\wedge$ and $\overline{\,\cdot\,}$.  We call the equation in the form of (\ref{lattice eq}) `lattice equation' for short.  Let us consider the initial value problem of (\ref{lattice eq}) from the initial data $\{u_j^0\}$ assuming $n=0$ is an initial time without loss of generality.  If we use the above equation recursively, we obtain the formal solution expressed by $u_j^0$ as
\begin{equation}
  u_j^n=f^{(n)}(u_{j-n}^0,u_{j-n+1}^0,\ldots,u_{j+n-1}^0,u_{j+n}^0),
\end{equation}
where $f^{(n)}$ is defined recursively by
\begin{eqnarray}
  & f^{(k+1)}(a_{-k-1},a_{-k},\ldots,a_k,a_{k+1}) \nonumber\\
= &f(f^{(k)}(a_{-k-1},\ldots,a_{k-1}),f^{(k)}(a_{-k},\ldots,a_k),f^{(k)}(a_{-k+1},\ldots,a_{k+1})).
\end{eqnarray}
However this solution contains $3^n$ terms of $u_j^0$ and it does not give any information on the solution since it is formal.  In this article, we show the list of evolution equations with a general solution containing the terms $u_j^0$ of which the number is of polynomial order.  If the solution $u_j^n$ can be expressed by the $O(n^m)$ terms among $\{u_j^0\}$, let us classify as the equation is of class $P_m$ (polynomial class of order $m$).\par
  For example, let us consider the equation
\begin{equation}  \label{example}
  u_j^{n+1}=u_{j-1}^n\wedge u_j^n\wedge u_{j+1}^n.
\end{equation}
Using the basic laws of lattice shown above, we can obtain a reduced expression of solution,
\begin{equation}  \label{solution example}
  u_j^n=u_{j-n}^0\wedge u_{j-n+1}^0\wedge\ldots\wedge u_{j+n-1}^0\wedge u_{j+n}^0.
\end{equation}
Equation (\ref{example}) is of class $P_1$ since the RHS of (\ref{solution example}) contains $2n+1$ ($=O(n)$) terms which is equal to the infimum of the terms from $u_{j-n}^0$ to $u_{j+n}^0$.\par
  There are equivalent equations through the transformation of variables and coordinates.  If $f_1(a,b,c)=f_2(c,b,a)$, then the solution to $u_j^{n+1}=f_1(u_{j-1}^n,u_j^n,u_{j+1}^n)$ is obtained from the solution to $u_j^{n+1}=f_2(u_{j-1}^n,u_j^n,u_{j+1}^n)$ through the transformation $j\to-j$.  Similarly, the following evolution rules given by $f_i$'s are equivalent each other,
\begin{eqnarray}
f_1(a,b,c)=f_2(c,b,a),         &\mbox{(reflection)} \\
f_1(a,b,c)=\overline{f_2(\overline{a},\overline{b},\overline{c})}, & \mbox{(conjugation)} \\
f_1(a,*,*)=f_2(*,b,*)=f_3(*,*,c), & \mbox{(Galilean transformation)} \\
f_1(a,b,*)=f_2(*,b,c),         & \mbox{(Galilean transformation)} \\
f_1(a,*,c)=f_2(a,b,*), & \mbox{\hskip-1cm(separation of even and odd sites)}
\end{eqnarray}
where the symbol `$*$' of arguments denotes $f_i$ does not depend on the corresponding argument.
Moreover if $f_1$ and $f_2$ satisfy the condition
\begin{equation}
  f_1(a,b,c)=\overline{f_2(a,\overline{b},c)}=\overline{f_2(\overline{a},b,\overline{c})},
\end{equation}
they are equivalent through the transformation $u_{2j}^n\to\overline{u_{2j}^n}$ and $u_{2j+1}^n\to u_{2j+1}^n$.  Similarly, if $f_1$ and $f_2$ satisfy the condition
\begin{equation}
  f_1(a,b,c)=f_2(\overline{a},\overline{b},\overline{c})=\overline{f_2(a,b,c)},
\end{equation}
they are equivalent through the transformation $u_j^{2n}\to\overline{u_j^{2n}}$ and $u_j^{2n+1}\to u_j^{2n+1}$.
\subsection{Elementary cellular automaton}
Elementary cellular automaton (ECA) is a simple evolutional digital system and the evolution rule is given by
\begin{equation}
  u_j^{n+1}=f_b(u_{j-1}^{n+1},u_j^n,u_{j+1}^n),
\end{equation}
where $j$ denotes an integer space site number, $n$ an integer time and $u$ a binary state value ($=0$ or 1)\cite{wolfram,wolfram2}.  The binary-valued function $f_b$ determines the time evolution rule and gives the state value at the next time.  Thus $f_b$ can be defined by a rule table in the following form,
\begin{equation}
\begin{array}{|c|c|c|c|c|c|c|c|c|}
\hline
  a\,b\,c  & 111 & 110 & 101 & 100 & 011 & 010 & 001 & 000 \\
\hline
  f_b(a,b,c) &  b_7 & b_6 & b_5 & b_4 & b_3 & b_2 & b_1 & b_0 \\
\hline
\end{array}
\end{equation}
where the upper row shows combinations of binary variables $a$, $b$, $c$ and the lower the binary values of $f(a,b,c)$.  Since $f$ is uniquely defined if all $b_k$'s are given, we can identify the evolution rule by the rule number $(r)_{10}=(b_7b_6\ldots b_1)_2$ after Wolfram\cite{wolfram,wolfram2}.  Let us call the ECA of the decimal rule number $r$ `ECA$r$'.  There exist 256 rules from ECA0 to ECA255.\par
Let us consider the lattice $(\mathbb{R},\le)$ and assume that the conjugation operator is given by
\begin{equation}
  \overline{a}=1-a.
\end{equation}
If we define the rule $f_b(a,b,c)$ using $\vee$, $\wedge$ and $\overline{\,\cdot\,}$, they are equivalent to the bitwise operations AND, OR and NOT respectively for the binary values as follows:
\begin{eqnarray}
  & a\vee b=\max(a,b)=a\,\mbox{OR}\,b,\qquad a\wedge b=\min(a,b)=a\,\mbox{AND}\,b,\nonumber\\
  &\overline{a}=1-a=\mbox{NOT}\,a,
\end{eqnarray}
where $a$, $b\in\{0,1\}$.  Therefore every ECA rule can be expressed by the lattice operations.\par
  Moreover, if we consider the lattice $(\mathbb{R},\le)$, (\ref{example}) becomes
\begin{equation}
  u_j^n=\min(u_{j-1}^n,u_j^n,u_{j+1}^n),
\end{equation}
and it gives ECA128 in the binary case.  The solution
\begin{equation}
  u_j^n=\min(u_{j-n}^0,u_{j-n+1}^0,\ldots,u_{j+n-1}^0,u_{j+n}^0),
\end{equation}
is 1 if $u_{j-n}^0=u_{j-n+1}^0=\ldots=u_{j+n-1}^0=u_{j+n}^0=1$ and 0 otherwise in the binary case.\par
  Considering the equivalent relations described in Subsection~\ref{subsec:evolution}, there exist 75 independent rules of ECA.  We list the rule numbers as follows:
\begin{eqnarray}
& \mbox{0, 1, 2, 3, 4, 6, 8, 9, 11, 12, 13, 14, 18, 19, 21, 22, 24, 25, 26, } \nonumber\\
& \mbox{27, 28, 29, 30, 32, 33, 35, 36, 37, 38, 40, 41, 42, 43, 44, 45,} \nonumber\\
& \mbox{46, 50, 54, 56, 57, 58, 60, 62, 72, 73, 74, 76, 78, 94, 104, 106,} \label{eca list} \\
& \mbox{108, 110, 122, 126, 128, 130, 132, 134, 136, 138, 140, 146,} \nonumber\\
& \mbox{150, 152, 154, 156, 162, 164, 168, 170, 172, 184, 200, 232.} \nonumber
\end{eqnarray}
\section{Solutions to evolution equations expressed by lattice operators}  \label{sec:solution}
In this section, we list up the lattice equations satisfying the following conditions.
\begin{itemize}
\item[(a)] Equation form obeys (\ref{lattice eq}).
\item[(b)] The terms $a$, $b$ and $c$ do not appear or appear once in $f(a,b,c)$.
\item[(c)] The solution is of polynomial class.
\end{itemize}
In addition, we show some equations which satisfy (a) and (c) but not (b) as an exceptional case.  All solutions are derived exactly by evaluating the initial value problem using lattice formulas shown in Subsection~\ref{subsec:lattice}.  Two examples of proofs on such solutions are shown in Appendix~\ref{sec:proof}.  In the list blow, we show the form of $f(a,b,c)$, the rule number of corresponding ECA and its solution.
\subsection{class $P_0$}
\begin{itemize}
\item
$f(a,b,c)=e$ ($\in L$) (ECA0)
\begin{equation*}
  u_j^n=e.
\end{equation*}
\item
$f(a,b,c)=a$ (ECA240)
\begin{equation*}
  u_j^n=u_{j-n}^0.
\end{equation*}
\item
$f(a,b,c)=\overline{a}\wedge b$ (ECA12)
\begin{equation*}
  u_j^n=\overline{u_{j-1}^0}\wedge u_j^0.
\end{equation*}
\item
$f(a,b,c)=\overline{a}\wedge\overline{b}$ (ECA3)
\begin{eqnarray*}
  & u_j^1=\overline{u_{j-1}^0}\wedge\overline{u_j^0},\quad u_j^2=u_{j-1}^0\vee(u_{j-2}^0\wedge u_j^0),\\
  & u_j^{2m+1}=u_{j-m}^1, \quad u_j^{2m+2}=u_{j-m}^2\quad (m\ge1).
\end{eqnarray*}
\item
$f(a,b,c)=\overline{a}\wedge\overline{b}\wedge c$ (ECA2)
\begin{equation*}
  u_j^n=\overline{u_{j+n-2}^0}\wedge\overline{u_{j+n-1}^0}\wedge u_{j+n}^0.
\end{equation*}
\item
$f(a,b,c)=\overline{a}\wedge b\wedge\overline{c}$ (ECA4)
\begin{equation*}
  u_j^n=\overline{u_{j-1}^0}\wedge u_j^0\wedge\overline{u_{j+1}^0}.
\end{equation*}
\item
$f(a,b,c)=\overline{a}\wedge\overline{b}\wedge\overline{c}$ (ECA1)
\begin{eqnarray*}
  & u_j^{2m-1}=\overline{u_{j-1}^0}\wedge\overline{u_j^0}\wedge\overline{u_{j+1}^0},\\ 
  & u_j^{2m}=(u_{j-2}^0\vee u_{j-1}^0\vee u_j^0)\wedge(u_{j-1}^0\vee u_j^0\vee u_{j+1}^0)\wedge(u_j^0\vee u_{j+1}^0\vee u_{j+2}^0).
\end{eqnarray*}
\item
$f(a,b,c)=(\overline{a}\vee\overline{b})\wedge c$ (ECA42)
\begin{equation*}
  u_j^n=(\overline{u_{j+n-2}^0}\vee\overline{u_{j+n-1}^0})\wedge u_{j+n}^0.
\end{equation*}
\item
$f(a,b,c)=(a\vee c)\wedge b$ (ECA200)
\begin{equation*}
  u_j^n=(u_{j-1}^0\vee u_{j+1}^0)\wedge u_j^0.
\end{equation*}
\item
$f(a,b,c)=(\overline{a}\vee\overline{c})\wedge b$ (ECA76)
\begin{equation*}
  u_j^n=(\overline{u_{j-1}^0}\vee\overline{u_{j+1}^0})\wedge u_j^0.
\end{equation*}
\item
$f(a,b,c)=(\overline{a}\vee\overline{c})\wedge\overline{b}$ (ECA19)
\begin{eqnarray*}
  u_j^1=&(\overline{u_{j-1}^0}\vee\overline{u_{j+1}^0})\wedge\overline{u_j^0}, \\
  u_j^{2m}=&(u_{j-1}^0\vee u_j^0)\wedge(u_j^0\vee u_{j+1}^0)\wedge(u_{j-2}^0\vee u_{j-1}^0\vee u_{j+1}^0\vee u_{j+2}^0), \\
  u_j^{2m+1}=&\overline{u_j^{2m}} \quad(m\ge1).
\end{eqnarray*}
\item
$f(a,b,c)=(b\vee(a\wedge c))\wedge(\overline{b}\vee(\overline{a}\wedge\overline{c}))$ (ECA36)
\begin{eqnarray*}
  u_j^1&=&(u_j^0\vee(u_{j-1}^0\wedge u_{j+1}^0))\wedge(\overline{u_j^0}\vee(\overline{u_{j-1}^0}\wedge\overline{u_{j+1}^0}), \\
  u_j^n&=&(u_j^0\vee(u_{j-1}^0\wedge u_{j+1}^0))\wedge(\overline{u_j^0}\vee(\overline{u_{j-1}^0}\wedge\overline{u_{j+1}^0}) \\
  &&\wedge(\overline{u_{j-1}^0}\vee u_j^0\vee\overline{u_{j+1}^0}\vee(u_{j-2}^0\wedge u_{j+2}^0)) \\
  &&\wedge(u_{j-1}^0\vee\overline{u_j^0}\vee u_{j+1}^0\vee(\overline{u_{j-2}^0}\wedge\overline{u_{j+2}^0})) \\
  &&\wedge(u_{j-1}^0\vee\overline{u_{j-1}^0}\vee u_{j+1}^0\vee\overline{u_{j+1}^0})) \quad (n\ge2).
\end{eqnarray*}
\end{itemize}
\subsection{class $P_1$}
We define new symbols about $\vee$ and $\wedge$ as
\begin{eqnarray*}
  & \Vee_{k_0\le k\le k_1}x_k=x_{k_0}\vee x_{k_0+1}\vee \cdots\vee x_{k_1}, \\
  & \Wedge_{k_0\le k\le k_1}x_k=x_{k_0}\wedge x_{k_0+1}\wedge \cdots\wedge x_{k_1}.
\end{eqnarray*}
\begin{itemize}
\item
$f(a,b,c)=a\wedge b$ (ECA192)
\begin{eqnarray*}
  u_j^n
&=\Wedge_{-n\le k\le 0}u_{j+k}^0 \\
&=u_{j-n}^0\wedge u_{j-n+1}^0\wedge\ldots\wedge u_j^0.
\end{eqnarray*}
\item
$f(a,b,c)=a\wedge b\wedge c$ (ECA128)
\begin{eqnarray*}
  u_j^n
&=\Wedge_{-n\le k\le n}u_{j+k}^0 \\
&=u_{j-n}^0\wedge u_{j-n+1}^0\wedge\ldots\wedge u_{j+n-1}^0\wedge u_{j+n}^0.
\end{eqnarray*}
\item
$f(a,b,c)=\overline{a}\wedge b\wedge c$ (ECA8)
\begin{eqnarray*}
  u_j^n&=\Bigl(\Wedge_{-1\le k\le n-2}\overline{u_{j+k}^0}\Bigr)\wedge\Bigl(\Wedge_{0\le k\le n}u_{j+k}^0\Bigr) \\
&=\overline{u_{j-1}^0}\wedge\overline{u_j^0}\wedge\ldots\wedge\overline{u_{j+n-2}^0}\wedge u_j^0\wedge u_{j+1}^0\wedge\ldots\wedge u_{j+n}^0.
\end{eqnarray*}
\item
$f(a,b,c)=a\wedge\overline{b}\wedge c$ (ECA32)
\begin{eqnarray*}
  u_j^n&=\Bigl(\Wedge_{0\le k\le n}u_{j-n+2k}^0\Bigr)\wedge\Bigl(\Wedge_{0\le k\le n-1}u_{j-n+k+1}^0\Bigr) \\
&=u_{j-n}^0\wedge\overline{u_{j-n+1}^0}\wedge u_{j-n+2}^0\wedge\overline{u_{j-n+3}^0}\wedge\ldots\wedge\overline{u_{j+n-1}^0}\wedge u_{j+n}^0.
\end{eqnarray*}
\end{itemize}
\subsection{class $P_2$}
\begin{itemize}
\item
$f(a,b,c)=(\overline{a}\vee b)\wedge c$ (ECA138)
\begin{eqnarray*}
  u_j^n
&=&\Wedge_{0\le l\le n}\Bigl(\Bigl(\Vee_{l-1\le k\le n-2}\overline{u_{j+k}^0}\Bigr)\vee u_{j+l}^0\Bigr) \\
&=&(\overline{u_{j-1}^0}\vee\overline{u_j^0}\vee\ldots\vee\overline{u_{j+n-2}^0}\vee u_j^0) \\
  &&\wedge(\overline{u_j^0}\vee\overline{u_{j+1}^0}\vee\ldots\vee\overline{u_{j+n-2}^0}\vee u_{j+1}^0) \\
  &&\ldots \\
  &&\wedge(\overline{u_{j+n-3}^0}\vee\overline{u_{j+n-2}^0}\vee u_{j+n-2}^0) \\
  &&\wedge(\overline{u_{j+n-2}^0}\vee u_{j+n-1}^0)\wedge u_{j+n}^0.
\end{eqnarray*}
\item
$f(a,b,c)=(a\vee\overline{b})\wedge c$ (ECA162)
\begin{eqnarray*}
  u_j^n
&=&\Wedge_{0\le l\le n}\Bigl(u_{j-n+2l}\vee\Bigl(\Vee_{l\le k\le n-1}\overline{u_{j-n+2k+1}^0}\Bigr)\Bigr) \\
&=&(u_{j-n}^0\vee\overline{u_{j-n+1}^0}\vee\overline{u_{j-n+3}^0}\vee\ldots\vee\overline{u_{j+n-3}^0}\vee\overline{u_{j+n-1}^0}) \\
  &&\wedge(u_{j-n+2}^0\vee\overline{u_{j-n+3}^0}\vee\overline{u_{j-n+5}^0}\vee\ldots\vee\overline{u_{j+n-3}^0}\vee\overline{u_{j+n-1}^0}) \\
  && \ldots \\
  &&\wedge(u_{j+n-4}^0\vee\overline{u_{j+n-3}^0}\vee\overline{u_{j+n-1}^0}) \\
  &&\wedge(u_{j+n-2}^0\vee\overline{u_{j+n-1}^0})\wedge u_{j+n}^0.
\end{eqnarray*}
\item
$f(a,b,c)=(\overline{a}\vee c)\wedge b$ (ECA140)
\begin{eqnarray*}
  u_j^n&=&\Wedge_{0\le l\le n}\Bigl(\Bigl(\Vee_{-1\le k\le n-l-2}u_{j+k}^0\Bigr)\vee u_{j+n-l}^0\Bigr) \\
&=&(\overline{u_{j-1}^0}\vee\overline{u_j^0}\vee\ldots\vee\overline{u_{j+n-3}^0}\vee\overline{u_{j+n-2}^0}\vee u_{j+n}^0) \\
  &&\wedge(\overline{u_{j-1}^0}\vee\overline{u_j^0}\vee\ldots\vee\overline{u_{j+n-3}^0}\vee u_{j+n-1}^0) \\
  &&\ldots \\
  &&\wedge(\overline{u_{j-1}^0}\vee\overline{u_j^0}\vee u_{j+2}^0) \\
  &&\wedge(\overline{u_{j-1}^0}\vee u_{j+1}^0)\wedge u_j^0.
\end{eqnarray*}
\item
$f(a,b,c)=(a\vee c)\wedge\overline{b}$ (ECA50)
\begin{eqnarray*}
  u_j^{2m-1}&=&\Bigl\{\Vee_{0\le l\le m-1}\Bigl(u_{j-2m+2l+1}^0\wedge\Bigl(\Wedge_{-m+l+1\le k\le m-l-1}\overline{u_{j+2k}^0}\Bigr)\Bigr)\Bigr\} \\
&&\vee\Bigl\{\Vee_{0\le l\le m-1}\Bigl(\Bigl(\Wedge_{-m+l+1\le k\le m-l-1}\overline{u_{j+2k}^0}\Bigr)\wedge u_{j+2m-2l-1}^0\Bigr)\Bigr\} \\
&&\vee\Bigl\{\Vee_{0\le l\le m-2}\Bigl(\overline{u_{j-2m+2l+2}^0}\wedge\Bigl(\Wedge_{-m+l+1\le k\le m-l-2}u_{j+2k+1}^0\Bigr)\Bigr)\Bigr\} \\
&&\vee\Bigl\{\Vee_{0\le l\le m-2}\Bigl(\Bigl(\Wedge_{-m+l+1\le k\le m-l-2}u_{j+2k+1}^0\Bigr)\wedge\overline{u_{j-2m+2l+1}^0}\Bigr)\Bigr\} \\
&=&(u_{j-2m+1}^0\wedge\overline{u_{j-2m+2}^0}\wedge\overline{u_{j-2m+4}^0}\wedge\ldots\wedge\overline{u_{j+2m-4}^0}\wedge\overline{u_{j+2m-2}^0}) \\
  &&\vee(\overline{u_{j-2m+2}^0}\wedge\overline{u_{j-2m+4}^0}\wedge\ldots\wedge\overline{u_{j+2m-4}^0}\wedge\overline{u_{j+2m-2}^0}\wedge u_{j+2m-1}^0) \\
  &&\vee(\overline{u_{j-2m+2}^0}\wedge u_{j-2m+3}^0\wedge u_{j-2m+5}^0\wedge\ldots\wedge u_{j+2m-5}^0\wedge u_{j+2m-3}^0) \\
  &&\vee(u_{j-2m+3}^0\wedge u_{j-2m+5}^0\wedge\ldots\wedge u_{j+2m-5}^0\wedge u_{j+2m-3}^0\wedge\overline{u_{j+2m-2}^0}) \\
  && \ldots \\
  &&\vee(\overline{u_{j-2}^0}\wedge u_{j-1}^0\wedge u_{j+1}^0)
   \vee(u_{j-1}^0\wedge u_{j+1}^0\wedge\overline{u_{j+2}^0}) \\
  &&\vee(u_{j-1}^0\wedge\overline{u_j^0})
  \vee(\overline{u_j^0}\wedge u_{j+1}^0), \\
  u_j^{2m}&=&\Bigl\{\Vee_{0\le l\le m}\Bigl(u_{j-2m+2l}^0\wedge\Bigl(\Wedge_{-m+l\le k\le m-l-1}\overline{u_{j+2k+1}^0}\Bigr)\Bigr)\Bigr\} \\
&&\vee\Bigl\{\Vee_{0\le l\le m}\Bigl(\Bigl(\Wedge_{-m+l\le k\le m-l-1}\overline{u_{j+2k+1}^0}\Bigr)\wedge u_{j+2m-2l}^0\Bigr)\Bigr\} \\
&&\vee\Bigl\{\Vee_{0\le l\le m-1}\Bigl(\overline{u_{j-2m+2l+1}^0}\wedge\Bigl(\Wedge_{-m+l+1\le k\le m-l-1}u_{j+2k}^0\Bigr)\Bigr)\Bigr\} \\
&&\vee\Bigl\{\Vee_{0\le l\le m-1}\Bigl(\Bigl(\Wedge_{-m+l+1\le k\le m-l-1}u_{j+2k}^0\Bigr)\wedge\overline{u_{j+2m-2l-1}^0}\Bigr)\Bigr\} \\
&=&(u_{j-2m}^0\wedge\overline{u_{j-2m+1}^0}\wedge\overline{u_{j-2m+3}^0}\wedge\ldots\wedge\overline{u_{j+2m-3}^0}\wedge\overline{u_{j+2m-1}^0}) \\
  &&\vee(\overline{u_{j-2m+1}^0}\wedge\overline{u_{j-2m+3}^0}\wedge\ldots\wedge\overline{u_{j+2m-3}^0}\wedge\overline{u_{j+2m-1}^0}\wedge u_{j+2m}^0) \\
  &&\vee(\overline{u_{j-2m+1}^0}\wedge u_{j-2m+2}^0\wedge u_{j-2m+4}^0\wedge\ldots\wedge u_{j+2m-4}^0\wedge u_{j+2m-2}^0) \\
  &&\vee(u_{j-2m+2}^0\wedge u_{j-2m+4}^0\wedge\ldots\wedge u_{j+2m-4}^0\wedge u_{j+2m-2}^0\wedge\overline{u_{j+2m-1}^0}) \\
  && \ldots \\
  &&\vee(u_{j-2}^0\wedge\overline{u_{j-1}^0}\wedge\overline{u_{j+1}^0})
  \vee(\overline{u_{j-1}^0}\wedge\overline{u_{j+1}^0}\wedge u_{j+2}^0) \\
  &&\vee(\overline{u_{j-1}^0}\wedge u_j^0)
  \vee(u_j^0\wedge\overline{u_{j+1}^0}). \\
\end{eqnarray*}
\item
$f(a,b,c)=(\overline{a}\vee\overline{b})\wedge\overline{c}$ (ECA21)
\begin{eqnarray*}
u_j^{2m-1}
&=& \Bigl\{\Vee_{0\le l\le m-2}\overline{u_{j+2l+1}^0}\Bigr\} \\
& & \wedge\Bigl\{\Wedge_{0\le l\le m-2}\Bigl(\overline{u_{j+l-1}^0}\vee\overline{u_{j+l}^0}\vee\Bigl(\Vee_{0\le k\le m-l-2}\overline{u_{j+2k+l+3}^0}\Bigr)\Bigr)\Bigr\} \\
& & \wedge\Bigl\{\Wedge_{0\le l\le m-2}\Bigl(\Bigl(\Vee_{0\le k\le m-l-2}\overline{u_{j+2k+l+2}^0}\Bigr)\vee\overline{u_{j+2m-l-1}^0}\Bigr)\Bigr\} \\
& & \wedge\Bigl\{\Wedge_{0\le l\le m-1}\Bigl(\overline{u_{j+l-1}^0}\vee\Bigl(\Vee_{0\le k\le m-l-1}\overline{u_{j+2k+l}^0}\Bigr)\Bigr)\Bigr\}, \\
u_j^{2m}
&=& \Bigl\{\Wedge_{0\le l\le m}u_{j+2l}^0\Bigr\} \\
& & \vee\Bigl\{u_{j-2}^0\wedge u_{j-1}^0\wedge\Bigl(\Wedge_{0\le l\le m-1}u_{j+2l+2}^0\Bigr\} \\
& & \vee\Bigl\{\Vee_{0\le l\le m-2}\Bigl(u_{j+l-1}^0\wedge u_{j+l}^0\wedge\Bigl(\Wedge_{0\le k\le m-l-2}u_{j+2k+l+3}^0\Bigr)\wedge u_{j+2m-l}^0\Bigr)\Bigr\} \\
& & \vee\Bigl\{\Vee_{0\le l\le m-2}\Bigl(u_{j+l}^0\wedge u_{j+l+1}^0\wedge\Bigl(\Wedge_{0\le k\le m-l-2}u_{j+2k+l+4}^0\Bigr)\Bigr)\Bigr\} \\
& & \vee\Bigl\{\Vee_{0\le l\le m-1}\Bigl(\Bigl(\Wedge_{0\le k\le m-l-1}u_{j+2k+l+1}^0\Bigr)\wedge u_{j+2m-l}^0\Bigr)\Bigr\} \\
& & \vee\Bigl\{\Vee_{0\le l\le m-1}\Bigl(u_{j+l}^0\wedge\Bigl(\Wedge_{0\le k\le m-l-1}u_{j+2k+l+1}^0\Bigr)\Bigr)\Bigr\}.
\end{eqnarray*}
\item
$f(a,b,c)=(\overline{a}\vee b)\wedge\overline{c}$ (ECA69)
\begin{eqnarray*}
u_j^{2m-1}&=(u_j^0\wedge\overline{u_{j+1}^0}) \\
&\vee\Bigl\{\Vee_{1\le l\le m-1}\Bigl(\Bigl(\Wedge_{0\le k\le l}\overline{u_{j+2k-1}^0}\Bigr)\wedge u_{j+2l}^0\Bigr)\Bigr\} \\
&\vee\Bigl\{\Vee_{1\le l\le m-1}\Bigl(\Bigl(\Wedge_{0\le k\le l}u_{j+2k}^0\Bigr)\wedge\overline{u_{j+2l+1}^0})\Bigr)\Bigr\} \\
&\vee\Bigl\{\Vee_{1\le l\le m-1}\Bigl(u_{j-2}^0\wedge\overline{u_{j-1}^0}\wedge\Bigl(\Wedge_{1\le k\le l}u_{j+2k}^0\Bigr)\wedge\overline{u_{j+2l+1}^0}\Bigr)\Bigr\} \\
&\vee\Bigl\{\Wedge_{0\le l\le m}\overline{u_{j+2l-1}^0}\Bigr\}, \\
u_j^{2m}&=(u_j^0\wedge\overline{u_{j+1}^0}) \\
&\vee\Bigl\{\Vee_{1\le l\le m}\Bigl(\Bigl(\Wedge_{0\le k\le l}\overline{u_{j+2k-1}^0}\Bigr)\wedge u_{j+2l}^0\Bigr)\Bigr\}\\
&\vee\Bigl\{\Vee_{1\le l\le m-1}\Bigl(\Bigl(\Wedge_{0\le k\le l}u_{j+2k}^0\Bigr)\wedge\overline{u_{j+2l+1}^0}\Bigr)\Bigr\} \\
&\vee\Bigl\{\Vee_{1\le l\le m-1}\Bigl(u_{j-2}^0\wedge\overline{u_{j-1}^0}\wedge\Bigl(\Wedge_{1\le k\le l}u_{j+2k}^0\Bigr)\wedge\overline{u_{j+2l+1}^0}\Bigr)\Bigr\} \\
&\vee\Bigl\{u_{j-2}^0\wedge\overline{u_{j-1}^0}\wedge\Bigl(\Wedge_{1\le l\le m}u_{j+2l}^0\Bigr)\Bigr\} \\
&\vee\Bigl\{\Wedge_{0\le l\le m}u_{j+2l}^0\Bigr\}.
\end{eqnarray*}
\item
$f(a,b,c)=(a\vee b)\wedge(b\vee c)\wedge(c\vee a)$ (ECA232)
\begin{eqnarray*}
  u_j^n&=&\Bigl\{\Wedge_{0\le l\le n-1}\Bigl(u_{j-l-1}^0\vee\Bigl(\Vee_{0\le k\le l}u_{j+2k-l}^0\Bigr)\Bigr)\Bigr\} \\
&&\wedge\Bigl\{\Wedge_{0\le l\le n-1}\Bigl(\Bigl(\Vee_{0\le k\le l}u_{j+2k-l}^0\Bigr)\vee u_{j+l+1}^0\Bigr)\Bigr\} \\
&&\wedge\Bigl\{\Vee_{0\le l\le n}u_{j+2l-n}^0\Bigr\} \\
&=&(u_{j-1}^0\vee u_j^0)\wedge(u_j^0\vee u_{j+1}^0) \\
        &&\wedge(u_{j-2}^0\vee u_{j-1}^0\vee u_{j+1}^0)\wedge(u_{j-1}^0\vee u_{j+1}^0\vee u_{j+2}^0) \\
        &&\wedge(u_{j-3}^0\vee u_{j-2}^0\vee u_j^0\vee u_{j+2}^0)\wedge(u_{j-2}^0\vee u_j^0\vee u_{j+2}^0\vee u_{j+3}^0) \\
        &&\ldots \\
        &&\wedge(u_{j-n}^0\vee u_{j-n+1}^0\vee u_{j-n+3}^0\vee\ldots\vee u_{j+n-1}^0) \\
        &&\qquad\wedge(u_{j-n+1}^0\vee u_{j-n+3}^0\vee\ldots\vee u_{j+n-1}^0\vee u_{j+n}^0) \\
        &&\wedge(u_{j-n}^0\vee u_{j-n+2}^0\vee\ldots\vee u_{j+n-2}^0\vee u_{j+n}^0).
\end{eqnarray*}
\item
$f(a,b,c)=b\wedge(\overline{a}\vee c)\wedge(a\vee\overline{c})$ (ECA132)
\begin{eqnarray*}
  u_j^n
&=&u_j^0\\
&&\wedge\Bigl\{\Wedge_{1\le l\le n}\Bigl(\Bigl(\Vee_{0\le k\le 2(l-1)}\overline{u_{j+k-l}^0}\Bigr)\vee u_{j+l}^0\Bigr)\Bigr\} \\
&&\wedge\Bigl\{\Wedge_{1\le l\le n}\Bigl(u_{j-l}^0\vee\Bigl(\Vee_{0\le k\le 2(l-1)}\overline{u_{j-k+l}^0}\Bigr)\Bigr)\Bigr\}
\\
&=&u_j^0\\
  &&\wedge(\overline{u_{j-1}^0}\vee u_{j+1}^0) \\
  &&\wedge(\overline{u_{j-2}^0}\vee\overline{u_{j-1}^0}\vee\overline{u_j^0}\vee u_{j+2}^0) \\
  &&\ldots \\
  &&\wedge(\overline{u_{j-n}^0}\vee\overline{u_{j-n+1}^0}\vee\ldots\vee\overline{u_{j+n-2}^0}\vee u_{j+n}^0) \\
  &&\wedge(u_{j-1}^0\vee\overline{u_{j+1}^0}) \\
  &&\wedge(u_{j-2}^0\vee\overline{u_j^0}\vee\overline{u_{j+1}^0}\vee\overline{u_{j+2}^0}) \\
  &&\ldots \\
  &&\wedge(u_{j-n}^0\vee\overline{u_{j-n+2}^0}\vee\overline{u_{j-n+3}^0}\vee\ldots\vee\overline{u_{j+n}^0}).
\end{eqnarray*}
\end{itemize}
\section{Asymptotic behavior of ECA}  \label{sec:asymptotic}
  For the solutions of polynomial class, we can easily grasp its asymptotic behavior for $n\gg0$ in the binary case.  We show it using two examples in Section~\ref{sec:solution}.\par
\begin{figure}[hbt]
\begin{center}
  \includegraphics[scale=1]{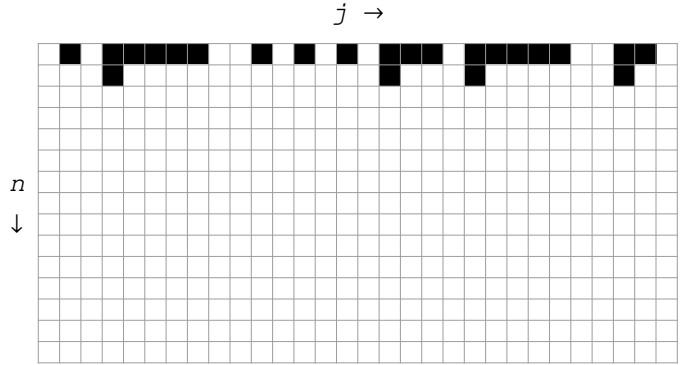}
\end{center}
  \caption{Evolution pattern of ECA8.  The symbols $\square$ and $\blacksquare$ denote 0 and 1 respectively.}
  \label{fig:eca8}
\end{figure}

  The evolution rule in the case of $f(a,b,c)=\overline{a}\wedge b\wedge c$ is equivalent to ECA8 in the binary case.  We show again the solution,
\begin{equation*}
  u_j^n=\overline{u_{j-1}^0}\wedge\overline{u_j^0}\wedge\ldots\wedge\overline{u_{j+n-2}^0}\wedge u_j^0\wedge u_{j+1}^0\wedge\ldots\wedge u_{j+n}^0.
\end{equation*}
In the binary case, $\overline{u}\wedge u=0$ since $u\in\{0,1\}$.  Thus the above solution is simplified as
\begin{eqnarray*}
  u_j^1&=\overline{u_{j-1}^0}\wedge u_j^0\wedge u_{j+1}^0
=
\cases{
  1 & ($u_{j-1}^0=0$ and $u_j^0=u_{j+1}^0=1$) \\
  0 & (otherwise)
}, \\
  u_j^n&=0 \quad (n\ge2).
\end{eqnarray*}
Figure~\ref{fig:eca8} shows a typical evolution pattern of ECA8.
\par
\begin{figure}[hbt]
\begin{center}
  \includegraphics[scale=1]{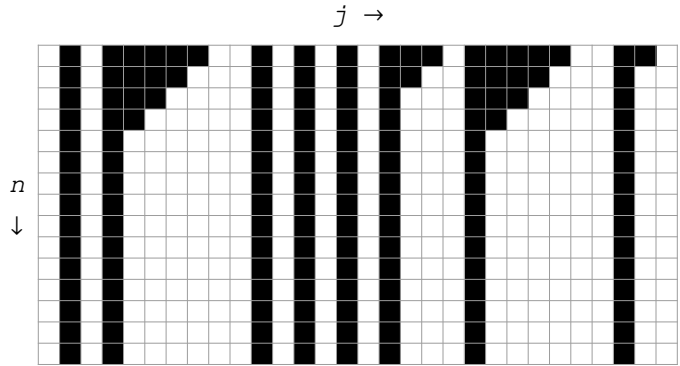}
\end{center}
  \caption{Evolution pattern of ECA140.}
  \label{fig:eca140}
\end{figure}

  Another example is $f(a,b,c)=(\overline{a}\vee c)\wedge b$ (ECA140).  The solution is
\begin{eqnarray}
  u_j^n=&(\overline{u_{j-1}^0}\vee\overline{u_j^0}\vee\ldots\vee\overline{u_{j+n-3}^0}\vee\overline{u_{j+n-2}^0}\vee u_{j+n}^0) \nonumber \\
  &\wedge(\overline{u_{j-1}^0}\vee\overline{u_j^0}\vee\ldots\vee\overline{u_{j+n-3}^0}\vee u_{j+n-1}^0) \nonumber \\
  &\ldots, \label{eca140} \\
  &\wedge(\overline{u_{j-1}^0}\vee\overline{u_j^0}\vee u_{j+2}^0) \nonumber \\
  &\wedge(\overline{u_{j-1}^0}\vee u_{j+1}^0)\wedge u_j^0. \nonumber
\end{eqnarray}
In the binary case, it is clear that $u_j^n=0$ ($n\ge1$) if $u_j^0=0$ and that $u_j^n=1$ ($n\ge1$) if $u_{j-1}^0=0$ and $u_j^0=1$.  Moreover, $u_j^n=0$ ($n\gg0$) if there exist $r\ge0$ such that $u_{j-1}^0=u_j^0=\cdots=u_{j+r}^0=1$ and $u_{j+r+1}^0=0$.  Therefore, we can see the asymptotic behavior of $u_j^n$ for large enough $n$ as
\begin{equation*}
  u_j^n=\cases{
1 & ($u_{j-1}^0=0$ and $u_j^0=1$) \\
0 & (otherwise)
}\quad (n\gg0)
\end{equation*}
Note that $u_j^n\equiv1$ ($n\ge1$) if $u_j^0\equiv1$ as a special case.  Figure~\ref{fig:eca140} shows a typical evolution pattern of ECA140.
\par
\begin{figure}[hbt]
\begin{center}
  \includegraphics[scale=1]{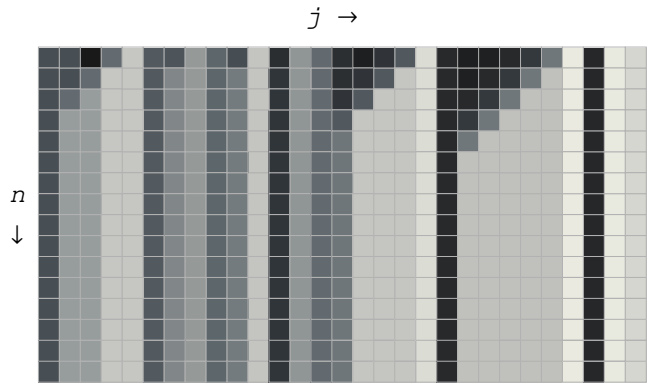}
\end{center}
  \caption{Evolution pattern of real valued solution.}
  \label{fig:eca140real}
\end{figure}

  Let us consider the asymptotic behavior of solution from real valued initial data for the same $f(a,b,c)=(\overline{a}\vee c)\wedge b$ of the lattice $(\mathbb{R},\le)$.  The lattice operations $a\vee b$, $a\wedge b$ and $\overline{a}$ can be replaced by $\max(a,b)$, $\min(a,b)$ and $1-a$ respectively for the lattice $(\mathbb{R},\le)$.  Thus (\ref{eca140}) can be rewritten as
\begin{eqnarray}
  u_j^n=\min(&\max(1-u_{j-1}^0,1-u_j^0,\ldots,1-u_{j+n-3}^0,1-u_{j+n-2}^0,u_{j+n}^0), \nonumber \\
  &\ldots, \nonumber \\
  &\underline{\max(1-u_{j-1}^0,1-u_j^0,\ldots,1-u_{j+r-2}^0,1-u_{j+r-1}^0,u_{j+r+1}^0)}, \nonumber \\
  &\underline{\max(1-u_{j-1}^0,1-u_j^0,\ldots,1-u_{j+r-2}^0,u_{j+r}^0)}, \label{eca140real2} \\
  &\ldots, \nonumber \\
  &\max(1-u_{j-1}^0,1-u_j^0,u_{j+2}^0), \nonumber \\
  &\max(1-u_{j-1}^0,u_{j+1}^0),u_j^0). \nonumber
\end{eqnarray}
If there exist $r\ge1$ such that $u_{j+r}^0<1/2$, the $\max$ terms above the underlined terms in (\ref{eca140real2}) are all greater than $\max(1-u_{j-1}^0,1-u_j^0,\ldots,1-u_{j+r-2}^0,u_{j+r}^0)$ since they include the same $1-u_{j-1}^0$, $1-u_j^0$, $\ldots$, $1-u_{j+r-2}^0$, and $1-u_{j+r}^0$ which is greater than $u_{j+r}^0$.  They can be neglected since RHS of (\ref{eca140real2}) is a minimum of all max terms.  Thus we obtain
\begin{eqnarray*}
  u_j^n=\min(
  &\max(1-u_{j-1}^0,1-u_j^0,\ldots,1-u_{j+r-2}^0,1-u_{j+r-1}^0,u_{j+r+1}^0), \\
  &\max(1-u_{j-1}^0,1-u_j^0,\ldots,1-u_{j+r-2}^0,u_{j+r}^0), \\
  &\ldots, \\
  &\max(1-u_{j-1}^0,1-u_j^0,u_{j+2}^0), \\
  &\max(1-u_{j-1}^0,u_{j+1}^0),u_j^0),
\end{eqnarray*}
and $u_j^{n+1}=u_j^n$.  Figure~\ref{fig:eca140real} shows an example of this type of solution.  Otherwise $u_k^0\ge1/2$ for any $k\ge j-1$.  In this case,
\begin{equation*}
  \max(1-u_{j-1}^0,1-u_j^0,\ldots,1-u_{j+r-2}^0,u_{j+r}^0)=u_{j+r}^0
\end{equation*}
and we have
\begin{equation*}
u_j^n=\min(u_j^0,u_{j+1}^0,\ldots,u_{j+n}^0),
\end{equation*}
from (\ref{eca140real2}). Then all $u$ become the same value $\min_{1\le k\le K}u_k^0$ for $n\gg0$ under the periodic boundary condition for the space cites with a finite period $K$.  In any case, the asymptotic behavior of solution is static ($u_j^{n+1}=u_j^n$) for $n\gg0$.
\par
  The behaviors of asymptotic solution shown in Figs.~\ref{fig:eca140} and \ref{fig:eca140real} are similar each other in a sense that they become static.  Though the binary solution is a special case of real valued solution, it implies that ECA can become a good test of the corresponding lattice equation.
\section{Concluding remarks}  \label{sec:conclusion}
We proposed the equations constructed from lattice operations $\vee$, $\wedge$ and $\overline{\,\cdot\,}$ of which solutions are expressed by the initial data of polynomial order.  Since ECA are embedded in these equations as a special case, we can grasp the asymptotic behavior of ECA easily using the solutions.\par
  We proposed 24 equations and their solutions.  Rule numbers of corresponding ECA are as follows:
\begin{quote}
0, 1, 2, 3, 4, 8, 12, 19, 21, 32, 36, 42, 50, 69, 76, 128, 132, 138, 140, 162, 192, 200, 232, 240.
\end{quote}
Therefore about one third of ECA listed in (\ref{eca list}) are solved.  ECA are roughly classified into 4 classes (Class 1 to 4) by Wolfram.  Almost ECA expressed by the lattice equation shown in this article are in Class 1 or 2, and some are in Class 3.  This classification is done by the geometrical complexity of solution pattern.  On the other hand, we classified the lattice equations from a viewpoint of algebraic complexity, that is, the order of initial data included in the solution.  It is interesting that some rule numbers in the same Wolfram class are not in the same polynomial class.\par
  The lattice operations in the lattice $(\mathbb{R},\le)$ can be embedded in the max-plus algebra where $\max$ ($\min$) is the addition, $+$ is the production, and $-$ is the subtraction\cite{gaubert}.  The max-plus equation
\begin{equation*}
  u_j^{n+1}=u_j^n+\max(u_{j-1}^n,1-u_j^n)-\max(u_j^n,1-u_{j+1}^n),
\end{equation*}
can not be expressed only by the lattice operations and is equivalent to ECA184 in the binary case\cite{nishinari}.  However, this equation is transformed into the lattice equation of $P_1$,
\begin{equation*}
  f_j^{n+1}=\max(f_{j-1}^n,f_{j+1}^n),
\end{equation*}
through the transformation
\begin{equation*}
  u_j^n=f_j^n-f_{j-1}^n+\frac{1}{2}.
\end{equation*}
There exist some max-plus equations which reduce to the lattice equation of polynomial class like the above example.  It is a future problem to clarify the relation between the max-plus and the lattice expressions from the viewpoint of solvability of initial data problem.\par
  We analyzed the lattice equations constructed from a simple combination of lattice operations.  Though they are simple, we can not confirm the polynomial class of some equations, for example, those defined by the following $f(a,b,c)$,
\begin{equation*}
\begin{array}{ll}
  (\overline{a}\vee c)\wedge \overline{b} \quad \mbox{(ECA35)}, & (a\vee b)\wedge c \quad \mbox{(ECA168)},\\
  (a\vee b)\wedge\overline{c} \quad \mbox{(ECA84)}, & (a\vee\overline{b})\wedge\overline{c} \quad \mbox{(ECA81)}.
\end{array}
\end{equation*}
They may not be in a polynomial class.  It is another future problem to evaluate the complexity of general solution to the above equations.  In addition, there exist many equations defined by the more complicated combination of lattice operations.  These propose a vast target to be analyzed.
\appendix
\section{Equivalent lattice equations and ECA rule numbers}  \label{sec:list}
In this appendix, we show in Table~\ref{table:equations} the equivalent lattice equations of polynomial class in the form,
\begin{equation}
  u_j^{n+1}=f(u_{j-1}^n,u_j^n,u_{j+1}^n),
\end{equation}
and its corresponding rule number of ECA.  The first column of each row shows equivalent $f(a,b,c)$'s and the first one is a representative.  The second column shows the class of its general solution.  The rule number of corresponding ECA is noted together with $f(a,b,c)$.  Note that classes of some lattice equations in the simple form described in Section~\ref{sec:solution} is not yet confirmed.  We show them in Table~\ref{table:unconfirmed}.
\begin{table}[hbt]
\begin{center}
\begin{tabular}{|l|l|}
\hline
$e \in L$ (0, 255) & $P_0$ \\
\hline
$a$ (240), $b$ (204), $c$ (170) , $\overline{a}$ (15), $\overline{b}$ (51), $\overline{c}$ (85) & $P_0$ \\
\hline
$a\wedge b$ (192), $b\wedge c$ (136), $a\wedge c$ (160), & $P_1$ \\
$a\vee b$ (152), $b\vee c$ (238), $a\vee c$ (250) & \\
\hline
$\overline{a}\wedge b$ (12), $\overline{b}\wedge c$ (34), $\overline{a}\vee b$ (207), $\overline{b}\vee c$, (187) & $P_0$ \\
$a\wedge\overline{b}$ (48), $b\wedge\overline{c}$ (68), $a\vee\overline{b}$ (243), $b\vee\overline{c}$ (221), & \\
$\overline{a}\wedge c$ (10), $\overline{a}\vee c$ (175), $a\wedge\overline{c}$ (80), $a\vee\overline{c}$ (245) & \\
\hline
$\overline{a}\wedge\overline{b}$ (3), $\overline{b}\wedge\overline{c}$ (17), 
$\overline{a}\vee\overline{b}$ (63), $\overline{b}\vee\overline{c}$ (119), & $P_0$ \\
$\overline{a}\wedge\overline{c}$ (5), $\overline{a}\vee\overline{c}$ (95) & \\
\hline
$a\wedge b\wedge c$ (128), $a\vee b\vee c$ (254) & $P_1$ \\
\hline
$\overline{a}\wedge b\wedge c$ (8), $a\wedge b\wedge \overline{c}$ (64), $\overline{a}\vee b\vee c$ (239), $a\vee b\vee \overline{c}$ (253) & $P_1$ \\
\hline
$a\wedge \overline{b}\wedge c$ (32), $a\vee \overline{b}\vee c$ (251) & $P_1$ \\
\hline
$\overline{a}\wedge \overline{b}\wedge c$ (2), $a\wedge \overline{b}\wedge \overline{c}$ (16), $\overline{a}\vee \overline{b}\vee c$ (191), $a\vee \overline{b}\vee \overline{c}$ (247) & $P_0$ \\
\hline
$\overline{a}\wedge b\wedge \overline{c}$ (4), $\overline{a}\vee b\vee \overline{c}$ (223) & $P_0$ \\
\hline
$\overline{a}\wedge \overline{b}\wedge \overline{c}$ (1), $\overline{a}\vee \overline{b}\vee \overline{c}$ (127) & $P_0$ \\
\hline
$(a\vee c)\wedge b$ (200), $(a\wedge c)\vee b$ (236) & $P_0$ \\
\hline
$(\overline{a}\vee c)\wedge b$ (140), $(a\vee \overline{c})\wedge b$ (196), 
$(\overline{a}\wedge c)\vee b$ (206), $(a\wedge \overline{c})\vee b$ (220) & $P_2$ \\
\hline
$(\overline{a}\vee \overline{c})\wedge b$ (76), $(\overline{a}\wedge \overline{c})\vee b$ (205) & $P_0$ \\
\hline
$(a\vee c)\wedge \overline{b}$ (50), $(a\wedge c)\vee \overline{b}$ (179) & $P_2$ \\
\hline
$(\overline{a}\vee \overline{c})\wedge \overline{b}$ (19), $(\overline{a}\wedge \overline{c})\vee \overline{b}$ (55) & $P_1$ \\
\hline
$(\overline{a}\vee b)\wedge c$ (138), $a\wedge(b\vee \overline{c})$ (208), 
$(\overline{a}\wedge b)\vee c$ (174), $a\vee(b\wedge \overline{c})$ (244) & $P_2$ \\
\hline
$(a\vee \overline{b})\wedge c$ (162), $a\wedge(\overline{b}\vee c)$ (176), 
$(a\wedge \overline{b})\vee c$ (186), $a\vee(\overline{b}\wedge c)$ (242) & $P_2$ \\
\hline
$(\overline{a}\vee \overline{b})\wedge c$ (42), $a\wedge(\overline{b}\vee \overline{c})$ (112), 
$(\overline{a}\wedge \overline{b})\vee c$ (171), $a\vee(\overline{b}\wedge \overline{c})$ (241) & $P_0$ \\
\hline
$(\overline{a}\vee b)\wedge \overline{c}$ (69), $\overline{a}\wedge(b\vee \overline{c})$ (13), 
$(\overline{a}\wedge b)\vee \overline{c}$ (93), $\overline{a}\vee(b\wedge \overline{c})$ (79) & $P_2$ \\
\hline
$(\overline{a}\vee \overline{b})\wedge \overline{c}$ (21), $\overline{a}\wedge(\overline{b}\vee \overline{c})$ (7), 
$(\overline{a}\wedge \overline{b})\vee \overline{c}$ (87), $\overline{a}\vee(\overline{b}\wedge \overline{c})$ (31) & $P_2$ \\
\hline
$(a\vee b)\wedge(b\vee c)\wedge(c\vee a)$ (232), $(\overline{a}\wedge\overline{b})\vee(\overline{b}\wedge\overline{c})\vee(\overline{c}\wedge\overline{a})$ (23) & $P_2$ \\
\hline
$b\wedge(\overline{a}\vee c)\wedge(a\vee\overline{c})$ (132), $\overline{b}\vee(\overline{a}\wedge c)\vee(a\wedge\overline{c})$ (123) & $P_2$ \\
\hline
$(b\vee(a\wedge c))\wedge(\overline{b}\vee(\overline{a}\wedge\overline{c}))$ (36), $(b\wedge(a\vee c))\vee(\overline{b}\wedge(\overline{a}\vee\overline{c}))$ (219)& $P_0$ \\
\hline
\end{tabular}
\end{center}
\caption{Equivalent lattice equations and their class.}
\label{table:equations}
\end{table}
\begin{table}[hbt]
\begin{center}
\begin{tabular}{|l|}
\hline
$(\overline{a}\vee c)\wedge \overline{b}$ (35), $(a\vee \overline{c})\wedge \overline{b}$ (49), 
$(\overline{a}\wedge c)\vee \overline{b}$ (59), $(a\wedge \overline{c})\vee \overline{b}$ (115) \\
\hline
$(a\vee b)\wedge \overline{c}$ (84), $\overline{a}\wedge(b\vee c)$ (14), 
$(a\wedge b)\vee \overline{c}$ (213), $\overline{a}\vee(b\wedge c)$ (143) \\
\hline
$(a\vee b)\wedge c$ (168), $a\wedge(b\vee c)$ (224), 
$(a\wedge b)\vee c$ (234), $a\vee(b\wedge c)$ (248) \\
\hline
$(a\vee \overline{b})\wedge \overline{c}$ (81), $\overline{a}\wedge(\overline{b}\vee c)$ (11), 
$(a\wedge \overline{b})\vee \overline{c}$ (117), $\overline{a}\vee(\overline{b}\wedge c)$ (47) \\
\hline
\end{tabular}
\end{center}
\caption{Equivalent lattice equations of unconfirmed class.}
\label{table:unconfirmed}
\end{table}
\section{Proof of solutions}  \label{sec:proof}
  The solutions described in the section \ref{sec:solution} are all derived only by using the formulas in the subsection \ref{subsec:lattice}.  In this appendix, we show two examples of proofs of solutions.  The solutions to the other equations can also be proved by the similar procedure.\par
  The first example is about the equation
\begin{equation}  \label{eca8}
  u_j^{n+1}=\overline{u_{j-1}^n}\wedge u_j^n\wedge u_{j+1}^n,
\end{equation}
which corresponds to ECA8.  The solution is
\begin{equation}  \label{eca8 sol}
  w_j^n=\overline{u_{j-1}^0}\wedge\overline{u_j^0}\wedge\ldots\wedge\overline{u_{j+n-2}^0}\wedge u_j^0\wedge u_{j+1}^0\wedge\ldots\wedge u_{j+n}^0.
\end{equation}
It is of class $P_1$.  For $n=1$, it is easy to check $w_j^1=\overline{u_{j-1}^0}\wedge u_j^0\wedge u_{j+1}^0$ from (\ref{eca8}) and (\ref{eca8 sol}).  If we substitute it into the RHS of (\ref{eca8}), we obtain
\begin{eqnarray*}
& & \overline{w_{j-1}^n}\wedge w_j^n\wedge w_{j+1}^n \\
&=& (\overline{\overline{u_{j-2}^0}\wedge\ldots\wedge\overline{u_{j+n-3}^0}\wedge u_{j-1}^0\wedge\ldots\wedge u_{j+n-1}^0}) \\
  && \wedge (\overline{u_{j-1}^0}\wedge\ldots\wedge\overline{u_{j+n-2}^0}\wedge u_j^0\wedge\ldots\wedge u_{j+n}^0) \\
& & \wedge (\overline{u_j^0}\wedge\ldots\wedge\overline{u_{j+n-1}^0}\wedge u_{j+1}^0\wedge\ldots\wedge u_{j+n+1}^0) \\
&=& (u_{j-2}^0\vee\ldots\vee u_{j+n-3}^0\vee\overline{u_{j-1}^0}\vee\ldots\vee\overline{u_{j+n-1}^0}) \\
& & \wedge (\overline{u_{j-1}^0}\wedge\ldots\wedge\overline{u_{j+n-1}^0}\wedge u_j^0\wedge\ldots\wedge u_{j+n+1}^0) \\
&=& \big((u_{j-2}^0\vee\ldots\vee u_{j+n-3}^0\vee\overline{u_{j-1}^0}\vee\ldots\vee\overline{u_{j+n-1}^0}) \\
& & \wedge \overline{u_{j-1}^0}\big)\wedge(\overline{u_j^0}\ldots\wedge\overline{u_{j+n-1}^0}\wedge u_j^0\wedge\ldots\wedge u_{j+n+1}^0) \\
&=& \overline{u_{j-1}^0}\wedge\overline{u_j^0}\ldots\wedge\overline{u_{j+n-1}^0}\wedge u_j^0\wedge\ldots\wedge u_{j+n+1}^0 \\
&=& w_j^{n+1}.
\end{eqnarray*}
Thus the induction on the solution holds.\par
  The second example is about the equation
\begin{equation}  \label{eca140-a}
  u_j^{n+1}=(\overline{u_{j-1}^n}\vee u_{j+1}^n)\wedge u_j^n,
\end{equation}
which corresponds to ECA140.  The solution is
\begin{eqnarray}  \label{eca140 sol}
  w_j^n
&=&(\overline{u_{j-1}^0}\vee\overline{u_j^0}\vee\ldots\vee\overline{u_{j+n-3}^0}\vee\overline{u_{j+n-2}^0}\vee u_{j+n}^0) \nonumber \\
  &&\wedge(\overline{u_{j-1}^0}\vee\overline{u_j^0}\vee\ldots\vee\overline{u_{j+n-3}^0}\vee u_{j+n-1}^0) \nonumber \\
  &&\ldots \\
  &&\wedge(\overline{u_{j-1}^0}\vee\overline{u_j^0}\vee u_{j+2}^0) \nonumber \\
  &&\wedge(\overline{u_{j-1}^0}\vee u_{j+1}^0)\wedge u_j^0. \nonumber 
\end{eqnarray}
It is more complicated than that of the first example and is of class $P_2$.  For $n=1$, it is easy to check $w_j^1=(\overline{u_{j-1}^0}\vee u_{j+1}^0)\wedge u_j^0$ from (\ref{eca140-a}) and (\ref{eca140 sol}).  We can rewrite $\overline{w_{j-1}^n}$, $w_j^n$ and $w_{j+1}^n$ as follows:
\begin{eqnarray*}
  \overline{w_{j-1}^n}
&=& \overline{(\overline{u_{j-2}^0}\vee\ldots\vee\overline{u_{j+n-3}^0}\vee u_{j+n-1}^0)\wedge\ldots\wedge(\overline{u_{j-2}^0}\vee u_j^0)\wedge u_{j-1}^0} \\
&=& \underbrace{(u_{j-2}^0\wedge\ldots\wedge u_{j+n-3}^0\wedge\overline{u_{j+n-1}^0})\vee\ldots\vee(u_{j-2}^0\wedge\overline{u_j^0})}_{A}\vee\underbrace{\overline{u_{j-1}^0}}_{B}, \\
  w_j^n
&=& (\overline{u_{j-1}^0}\vee\ldots\vee\overline{u_{j+n-2}^0}\vee u_{j+n}^0)\wedge\ldots\wedge(\overline{u_{j-1}^0}\vee u_{j+1}^0)\wedge u_j^0 \\
&=& \Big(\underbrace{\overline{u_{j-1}^0}}_{B}\vee\Big(\underbrace{(\overline{u_j^0}\vee\ldots\vee\overline{u_{j+n-2}^0}\vee u_{j+n}^0)\wedge\ldots\wedge(\overline{u_j^0}\vee u_{j+2}^0)\wedge u_{j+1}^0}_{C}\Big)\Big) \\
&& \wedge\underbrace{u_j^0}_{D}, \\
  w_{j+1}^n
&=& (\overline{u_j^0}\vee\ldots\vee\overline{u_{j+n-1}^0}\vee u_{j+n+1}^0)\wedge\ldots\wedge(\overline{u_j^0}\vee u_{j+2}^0)\wedge u_{j+1}^0 \\
&=& \underbrace{(\overline{u_j^0}\vee\ldots\vee\overline{u_{j+n-1}^0}\vee u_{j+n+1}^0)}_{E} \\
& & \qquad\wedge\Big(\underbrace{(\overline{u_j^0}\vee\ldots\vee\overline{u_{j+n-2}^0}\vee u_{j+n}^0)\wedge\ldots\wedge(\overline{u_j^0}\vee u_{j+2}^0)\wedge u_{j+1}^0}_{C}\Big).
\end{eqnarray*}
If we substitute (\ref{eca140 sol}) into the RHS of (\ref{eca140-a}), we obtain
\begin{eqnarray*}
  (\overline{w_{j-1}^n}\vee w_{j+1}^n)\wedge w_j^n
&=& ((A\vee B)\vee(E\wedge C))\wedge((B\vee C)\wedge D) \\
&=& (A\vee B\vee E)\wedge(A\vee B\vee C)\wedge(B\vee C)\wedge D \\
&=& (A\vee B\vee E)\wedge(B\vee C)\wedge D \\
\end{eqnarray*}
Moreover, we have
\begin{eqnarray*}
  A\vee E
&=& \underbrace{(u_{j-2}^0\wedge\ldots\wedge u_{j+n-3}^0\wedge\overline{u_{j+n-1}^0})\vee\ldots\vee(u_{j-2}^0\wedge\overline{u_j^0})}_{A} \\
& & \qquad\vee
\underbrace{(\overline{u_j^0}\vee\ldots\vee\overline{u_{j+n-1}^0}\vee u_{j+n+1}^0)}_{E} \\
&=& \Bigl((u_{j-2}^0\wedge\ldots\wedge u_{j+n-3}^0\wedge\overline{u_{j+n-1}^0})\vee\overline{u_{j+n-1}^0}\Bigr) \\
& & \qquad\vee\ldots\vee
  \Bigl((u_{j-2}^0\wedge\overline{u_j^0})\vee\overline{u_j^0}\Bigr)\vee u_{j+n+1}^0 \\
&=& \overline{u_{j+n-1}^0}\vee\ldots\vee\overline{u_j^0}\vee u_{j+n+1}^0 \\
&=& E.
\end{eqnarray*}
Therefore,
\begin{eqnarray*}
&&  (\overline{w_{j-1}^n}\vee w_{j+1}^n)\wedge w_j^n \\
&=& (B\vee E)\wedge(B\vee C)\wedge D \\
&=& (\overline{u_{j-1}^0}\vee\overline{u_j^0}\vee\ldots\vee\overline{u_{j+n-1}^0}\vee u_{j+n+1}^0) \\
& & \wedge(\overline{u_{j-1}^0}\vee\ldots\vee\overline{u_{j+n-2}^0}\vee u_{j+n}^0)\wedge\ldots\wedge(\overline{u_{j-1}^0}\vee u_{j+1}^0)\wedge u_j^0 \\
&=& w_j^{n+1},
\end{eqnarray*}
is derived and the induction on the solution holds.
\section*{References}


\begin{thebibliography}{99}
\bibitem{gratzer} Gr\"atzer G 1971 {\it Lattice Theory: First Concepts and Distributive Lattices} (San Francisco: W. H. Freeman)
\bibitem{bellon} Bellon M P and Viallet C M 1999 {\it Commun. Math. Phys.} {\bf 204} 425--37
\bibitem{grammaticos} Grammaticos B, Ramani A and Viallet C M 2005 {\it Physics Letters A} {\bf 336} 152--8
\bibitem{gaubert} Gaubert S and Plus M 1997 {\it Lecture Notes in Computer Science} {\bf 1200} 261--82
\bibitem{wolfram}
Wolfram S 1986 {\it Theory and Applications of Cellular Automata} (Singapore: World Scientific)
\bibitem{wolfram2}
Wolfram S 2002 {\it A New Kind of Science} (Champaign: Wolfram Media)
\bibitem{takahashi}
Takahashi D and Satsuma J 1990 {\it J. Phys. Soc. Jpn.} {\bf 59} 3514--9
\bibitem{tokihiro}
Tokihiro T, Takahashi D, Matsukidaira J and Satsuma J 1996 {\it Phys. Rev. Lett.} {\bf 76} 3247--50
\bibitem{nishinari}
Nishinari K and Takahashi D 1998 {\it J. Phys. A} {\bf 31} 5439--50
\bibitem{takahashi2}
Takahashi D, Matsukidaira J, Hara H and Feng B 2011 {\it J. Phys. A: Math. Theor.} {\bf 44} 135102 (21pp)
\end{thebibliography}
\end{document}